\documentclass[prl,twocolumn,showpacs,floatfix]{revtex4-2}
\pdfoutput=1
\usepackage[usenames,dvipsnames]{color}
\usepackage{longtable}
\usepackage{epsfig}
\usepackage{amssymb}
\usepackage{amsmath}
\usepackage{verbatim}
\usepackage{xspace}
\usepackage{titlesec}
\usepackage{babel}
\usepackage{float}
\titleformat{\section}{\normalfont\fontsize{10}{15}\bfseries\filcenter}{\thesection}{1em}{}

\definecolor{darkblue}{rgb}{0,0,0.5}
\definecolor{darkgreen}{rgb}{0.1,0,0.3}
\definecolor{darkred}{rgb}{0.6,0,0}
\usepackage{graphics,hyperref}

\newcommand{\ba}{\begin{eqnarray}}
\newcommand{\ea}{\end{eqnarray}}
\newcommand{\be}{\begin{equation}}
\newcommand{\ee}{\end{equation}}

\begin{document}
\title{Ghost-Induced Phase Transition in the Final Stages of Black Hole Evaporation}
\author{A.~Bonanno${}^{1,2}$ and S. Silveravalle${}^{3,4,5}$}

\affiliation{
\mbox{${}^1$INAF, Osservatorio Astrofisico di Catania, Via S. Sofia 78, 95123 Catania, Italy}
\mbox{${}^2$INFN, Sezione di Catania, Via S. Sofia 64, 95123 Catania, Italy}
\mbox{${}^3$SISSA - International School for Advanced Studies, Via Bonomea 265, 34136 Trieste, Italy}
\mbox{${}^4$INFN, Sezione di Trieste, Via Valerio 2, 34127 Trieste, Italy}
\mbox{${}^5$IFPU - Institute for Fundamental Physics of the Universe, Via Beirut 2, 34151 Trieste, Italy}
}

\begin{abstract}
We explore a novel scenario in which a quantum-induced ghost instability drives the natural evolution of an evaporating Schwarzschild black hole toward a stable naked singularity. This process, arising from quadratic curvature corrections to the Einstein-Hilbert action at high energies, circumvents the inconsistencies associated with classical naked singularities. The onset of ghost-driven instability signals a phase transition that fundamentally alters black hole evaporation, rendering the information paradox moot as it merges with the singularity issue. Our findings suggest a new pathway for black hole evolution at high-energy scales, offering insights that may bridge key gaps until a full theory of quantum gravity is realized.
\end{abstract}

\maketitle

\section{Introduction}

Black holes are among the simplest yet most distinctive and fascinating predictions of General Relativity. Despite their simplicity, they encapsulate some of the fundamental inconsistencies within Einstein’s theory. Internally they exhibit singularities where spacetime ceases to be well defined, while externally they prevent quantum fields from following a unitary evolution. Both of these issues have numerous proposed solutions \cite{Bambi:2023try,Chen:2014jwq,Page:1979tc}, but a consensus is still lacking. While the first issue—the singularity problem—may be temporarily set aside by invoking the cosmic censorship conjecture, which posits that singularities are hidden from observation \cite{Penrose:1969pc,Wald:1997wa}, the second issue—the information paradox—presents in principle an observable phenomenon.
Hawking, with a solid framework for describing the quantum vacuum in the presence of an event horizon, demonstrated that black holes emit radiation \cite{Hawking:1975vcx}. The dynamics of their evaporation is derived from the equations of motion and, in the context of General Relativity, it predicts that in the final stages they emit radiation at extremely high temperatures, ultimately disappearing and leaving behind Minkowski space filled with thermal radiation \cite{Hawking:1974rv}.
The information paradox can be illustrated by considering a star in a known microstate that collapses into a black hole. After the collapse, information about the initial microstate becomes inaccessible, and the black hole is indistinguishable from another with the same mass and angular momentum. The prevailing view is that the information is not lost at this stage, but rather hidden behind the event horizon. However, if the black hole undergoes complete evaporation, the remaining thermal radiation encodes only information about the external spacetime: while information is now fully accessible, the knowledge of the original microstate is irrevocably lost. 

The root of the information paradox, however, lies in the singularity problem, as the loss of information occurs at the moment the singularity vanishes. It is widely expected that both of these issues will ultimately be resolved by a quantum theory of gravity. The small spatial scales associated with the singularity and the extreme energies present during the final stages of black hole evaporation suggest that quantum effects must be incorporated into the gravitational sector of the theory.
In this paper, we do not seek to resolve these issues fully or to provide a complete quantum description of gravity. Instead, we aim to explore the effects of the first quantum corrections on the evolution of an evaporating black hole. We argue that a quantum-corrected, though intrinsically incomplete, theory of gravity can lead to a dynamical phase transition that results in a naked singularity at the end of evaporation. This scenario effectively merges the singularity problem with the information paradox, presenting a unified perspective on both challenges.

\section{Quadratic gravity and its black hole solutions}

Perturbative \cite{tHooft:1974toh}, non-perturbative \cite{Benedetti:2013jk} and fundamental \cite{Zwiebach:1985uq} approaches to quantum gravity converge on the idea that quadratic curvature terms represent the first quantum corrections to General Relativity. The most general quadratic action can be written as
\begin{equation}\label{action}
\mathcal{S}=\int\mathrm{d}^4x\sqrt{-g}\left[\gamma R-\alpha C^{\mu\nu\rho\sigma}C_{\mu\nu\rho\sigma}+\beta R^2+\eta\mathcal{G}\right],
\end{equation}
where $\mathcal{G}$ is the topological Gauss-Bonnet combination, $R$ is the Ricci scalar, $C_{\mu\nu\rho\sigma}$ is the Weyl tensor, $\gamma$=$1/16\pi\,G$ and $\alpha$, $\beta$, $\eta$ are dimensionless parameters. The analysis of perturbations shows the presence of three modes: the standard massless tensor mode, a massive scalar mode of mass $m_0=\sqrt{\gamma/6\beta}$ and a massive tensor mode of mass $m_2=\sqrt{\gamma/2\alpha}$. At the quantum level this theory is renormalizable but non-unitary due to the negative sign in front of the propagator of the massive tensor mode \cite{Stelle:1976gc}; this quantum ghost is indeed the Ostrogradsky ghost expected by the higher derivatives present in the theory, and is usually considered as the proof of the incompleteness of the theory. Nonetheless, classical solutions of quadratic theories of gravity, in particular Starobinsky model of inflation \cite{Starobinsky:1980te}, agree well with observational data \cite{Planck:2018vyg}.

\begin{figure}[htb]
\includegraphics[width=\columnwidth]{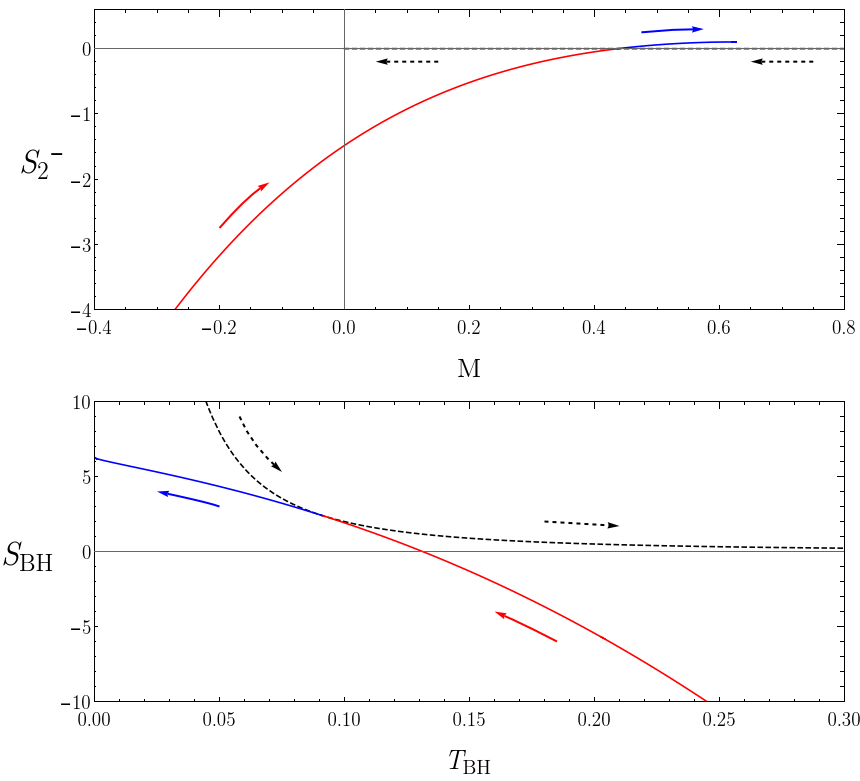}
\caption{Gravitational and thermodynamical properties of black holes in quadratic gravity in units of $m_2$; the dashed line indicates Schwarzschild black holes and the solid lines non-Schwarzschild ones with a blue line for Yukawa repulsive and a red one for Yukawa attractive solutions. The arrows indicate the direction of decreasing horizon radius.}
\label{fig:metric}
\end{figure}

Static, spherically symmetric solutions of \eqref{action} exhibit a greater variety than those in General Relativity, with three families of exotic solutions, Schwarzschild and non-Schwarzschild black holes \cite{Silveravalle:2022lid,Daas:2022iid,Silveravalle:2023lnl}. However, black holes are guaranteed to have a vanishing Ricci scalar \cite{Nelson:2010ig,Lu:2015cqa}, rendering the Ricci scalar squared term trivial. Thus, they can be studied within Einstein-Weyl theory \cite{Silveravalle:2022wij}. In the Newtonian limit, their gravitational potential is
\begin{equation}\label{potential}
    \Phi_N(r)=-\frac{M}{r}+S_2^-\frac{\mathrm{e}^{-m_2\,r}}{r},
\end{equation}
where $M$ represents the total mass, $S_2^-$ is a free parameter which we will call Yukawa charge, and together they completely characterize the field at large distances. The behavior of the metric near the horizon defines the thermodynamical properties through Hawking’s definition of temperature \cite{Hawking:1975vcx} and Wald’s definition of entropy \cite{Wald:1993nt}. The properties of black holes in quadratic gravity are shown in Fig. \ref{fig:metric}.

Non-Schwarzschild black holes can have either positive or negative Yukawa charges, leading to either repulsive or attractive contributions from the Yukawa term \cite{Lu:2015psa,Goldstein:2017rxn,Bonanno:2019rsq}.
The branches of Schwarzschild and non-Schwarzschild black holes cross each other at a specific mass $M_c$, which discriminates non-Schwarzschild black holes in Yukawa attractive and Yukawa repulsive. Yukawa-attractive black holes exhibit unbounded negative mass and entropy, as well as large temperature and horizon radius; Yukawa-repulsive black holes, on the other hand, have a limited range for mass and entropy (which are always positive) and smaller temperature and horizon radius. An analysis of the metric near the origin reveals that the metric of Yukawa attractive black holes diverges as $g_{tt}\sim r^{-1}$ and $g^{rr}\sim r^{-1}$, while for Yukawa repulsive black holes vanishes as $g_{tt}\sim r^{2}$ and $g^{rr}\sim r^{-2}$ \cite{Bonanno:2019rsq}. Finally, the analysis of linear perturbations shows that Schwarzschild black holes with mass $M<M_c$ and Yukawa repulsive black holes are unstable due to an exponential growth of the massive tensor mode of perturbations, while large Schwarzschild black holes and Yukawa attractive black holes are stable \cite{Lu:2017kzi,Held:2022abx,Konoplya:2025afm}. 

To gain insight into the role of non-Schwarzschild black holes in our universe, it is crucial to examine the crossing point with physical units. As shown in \cite{Bonanno:2024fcv}, the precise value depends on the loosely constrained $\alpha$ parameter in \eqref{action}. Rigorously, it can only be considered $M_c < 4.8 \times 10^{-8} M_\odot$ by employing constraints on Yukawa corrections to Newton potential in torsion balance experiments \cite{Giacchini:2016nta}; using an $\alpha$ value of the same order as the Starobinsky inflation parameter, we instead obtain a value of $M_c \sim 4.8 \times 10^{-34} M_\odot$. Such a small mass is clearly unattainable through astrophysical processes and can only be reached during the final stages of evaporation.

\section{Ghosts, instabilities and phase transitions}

In a Ricci scalar-flat spacetime, the modes of linear perturbations will satisfy the equations
\begin{equation}\label{eq:pert}
    \begin{split}
        \delta G_{\mu\nu}+m_2^2\left(\delta f_{\mu\nu}-\bar{g}_{\mu\nu}\delta f\right)&\,=0,\\
        \bar{\Box}\delta\psi+m_0^2\delta\psi &\,=0,\\
        \bar{\Box}\delta f_{\mu\nu}-\bar{\nabla}_\mu\bar{\nabla}_\nu\delta f+2\bar{R}_{\mu\rho\nu\sigma}\delta f^{\rho\sigma}+&\\
        +2\bar{f}^{\rho\sigma}\delta R_{\mu\rho\nu\sigma}-m_2^2\bigg(\delta f_{\mu\nu}-\bar{g}_{\mu\nu}\bar{f}^{\rho\sigma}\delta f_{\rho\sigma}+&\\
        -\left(\bar{g}_{\mu\nu}+\bar{f}_{\mu\nu}\right)\delta f-\frac{1}{2}\bar{f}^{\rho\sigma}\bar{f}_{\rho\sigma}\delta g_{\mu\nu}\bigg)&\,=0,
    \end{split}
\end{equation}
where $\delta g_{\mu\nu}$ is the massless tensor mode, $\delta\psi$ is the massive scalar mode, $\delta f_{\mu\nu}$ is the massive tensor mode and barred quantities refer to the background spacetime with $\bar{\psi}=\bar{f}=\bar{R}=0$ and $\bar{f}_{\mu\nu}=-m_2^{-2}\bar{R}_{\mu\nu}$ \cite{Held:2022abx}. The massive scalar mode is decoupled from the others and therefore behaves as a test field, contributing trivially to the dynamics. The massive tensor mode, on the other hand, is coupled to the metric perturbation if and only if $\bar{f}_{\mu\nu}=-m_2^{-2}\bar{R}_{\mu\nu}\neq 0$; otherwise, it simply acts as a source. This latter relation has a crucial consequence for the physical interpretation of the theory. The presence of a quantum ghost can lead to vacuum instability due to pair creation of ghost and non-ghost particles of total zero energy. Nonetheless, this applies only in the case where ghosts and standard particles are coupled, which, in our context, means only in a non-Schwarzschild background. The Hawking process, in which all possible modes—including massive tensor ghosts—are produced, cannot induce a vacuum instability for large Schwarzschild black holes; given that generally tensor modes \cite{Page:1976df} and massive particles \cite{Page:1977um} contribute with a small fraction of the total emission, we can conclude that an astrophysical or primordial black hole with mass $M_0$ will initially evaporate as predicted by General Relativity. 
However, at a fraction $t_c/t_e=1-M_c^3/M_0^3$ of the time it would take to fully evaporate, it reaches the crossing point with mass $M_c$ where it becomes unstable due to the behavior of the massive tensor mode. 

\begin{figure}[htb]
\includegraphics[width=\columnwidth]{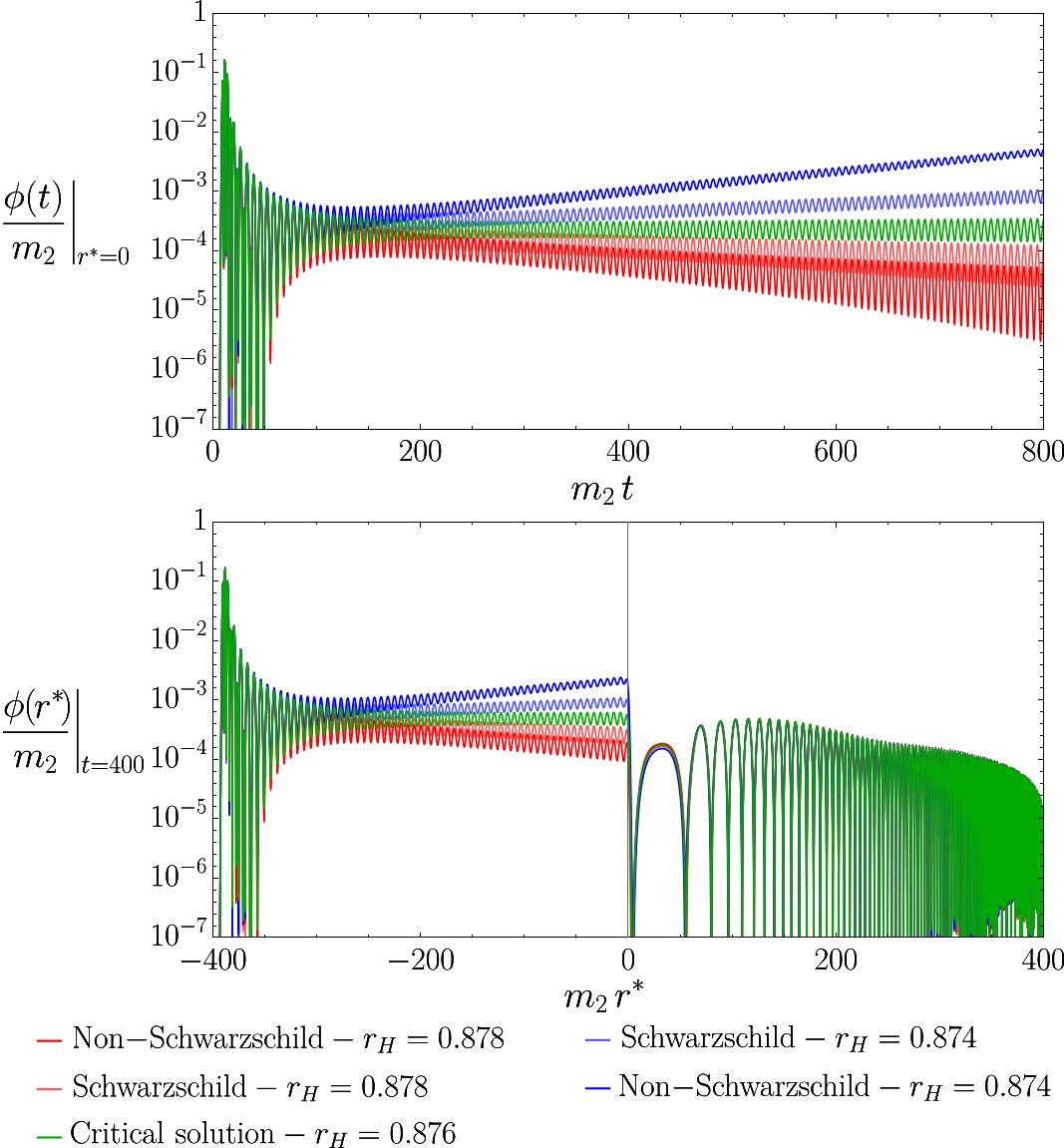}
\caption{Perturbations of black holes near the critical point: the top panel shows the time evolution, and the bottom panel displays the radial distribution of the perturbations.}
\label{fig:pert}
\end{figure}

For a quantitative analysis of the evolution of perturbations, we can rewrite equations \eqref{eq:pert} as a single wave equation for a function $\phi$, with a potential that is a complicated function of the background metric \cite{Held:2022abx}. We then numerically integrate the full wave equation to extract the time evolution and radial distribution of linear perturbations, as recently done in \cite{Konoplya:2025afm}. In Fig.~\ref{fig:pert}, we illustrate the linear perturbations for solutions near and at the crossing point.  The top panel shows that larger black holes are stable, with perturbations undergoing exponential suppression over time. In contrast, smaller black holes are subject to an instability, with their perturbations being exponentially amplified. A similar trend is observed in the radial distribution near the horizon, approximately for negative values of the tortoise coordinate.

For the crossing point solution, the perturbations display a constant oscillation profile in both the temporal and radial domains. Near the horizon, the perturbations of stable black holes follow the behavior:
\begin{equation}\label{perprof}
\phi(r,t) = \phi_{bo}(r,t)\,\mathrm{e}^{-\frac{t}{\tau}}\mathrm{e}^{-\frac{r^*}{\xi}},
\end{equation}
where $\phi_{bo}(r,t)$ is a bounded oscillatory function. Consequently, the limits $\tau \rightarrow \infty$ and $\xi \rightarrow \infty$ correspond to a critical point, signaling the divergence of the correlation length for an ensemble of perturbations at this transition, as first argued in \cite{Bonanno:2024fcv}.

To elucidate this transition, we note that the massive tensor modes driving it are perturbations of the Ricci tensor \cite{Held:2022abx}. At the critical point, these perturbations grow in magnitude, inducing a global change in the Ricci tensor. Since the Ricci tensor vanishes for Schwarzschild black holes but is non-zero for non-Schwarzschild solutions, it serves as the natural order parameter for this transition.

The associated ``symmetry breaking'' is evident in the behavior of geodesic congruences. In Ricci-flat spacetimes, geodesics focus, defocus, or remain parallel solely due to their initial conditions. This is not the case in Ricci-curved spacetimes, where the $R_{\mu\nu}x^\mu x^\nu$ term in the Raychaudhuri equation forces geodesics to focus (or defocus) if the Ricci tensor along their trajectory $x^\mu$ is positive (or negative). Specifically, for a family of causal geodesics with energy $E$, the Raychaudhuri scalar at large distances takes the form:
\begin{equation}\label{foc}
    R_{\mu\nu}x^\mu x^\nu \sim S_2^- \frac{E^2 m_2^2\,\mathrm{e}^{-m_2\,r}}{r}.
\end{equation}
The transition to the Yukawa attractive or repulsive branches thus corresponds to the focusing or defocusing of initially parallel geodesics, which can be considered as a sort of order parameter in this context.

However, Yukawa attractive and repulsive black holes represent distinct physical objects, and it is crucial to determine which branch the phase transition selects. In \cite{Bonanno:2024fcv}, we showed that the evaporation process in the Yukawa-attractive branch leads to unphysical predictions, without any compelling physical mechanism to exclude this scenario. Here, we address this challenge through the lens of phase transitions.

A key distinguishing feature lies in the linear stability of the solutions. Yukawa-attractive black holes are stable, maintaining dynamical equilibrium with their environment, which makes the phase transition an equilibrium process. In contrast, Yukawa-repulsive black holes are unstable, resulting in a dynamical, non-equilibrium phase transition.

When random fluctuations lack a preferred sign, they may infinitesimally perturb the critical solution toward both branches of non-Schwarzschild black holes. However, while a Yukawa-attractive black hole would require the slow timescale of evaporation to significantly deviate from a Schwarzschild solution (with random fluctuations potentially restoring it to the critical point), a Yukawa-repulsive black hole experiences an exponentially growing mode that rapidly drives it away from the Schwarzschild branch. 
This reveals an important principle: when three phases coexist at a critical point but one phase is unstable, the system itself becomes unstable and evolves along the direction of instability.

\section{Physical insight into the non-linear evolution}

While a complete description of the transition needs a non-linear analysis of the time-dependent equations, some key physical aspects can be extracted from their approximations. 
The general metric can be written as
\begin{equation}\label{metric}
    \mathrm{d}s^2=-\mathrm{e}^{\tau(r,t)}\mathrm{d}t^2+\mathrm{e}^{-\sigma(r,t)}\mathrm{d}r^2+r^2\mathrm{d}\Omega^2;
\end{equation}
close to the origin the field equations simplify and reduce to the autonomous dynamical system 
\begin{equation}\label{linsys}
\begin{split}
    \frac{\partial\tau}{\partial x}=&\,\frac{1}{2}\left(4+2\tau+4\sigma+\tau^2+\tau\sigma\right),\\
    \frac{\partial\sigma}{\partial x}=&\,\frac{8+8\sigma-\sigma^2+3\tau^2+\sigma\tau^2+2\sigma^2\tau-\tau^3}{2\left(\tau-2\right)},
\end{split}
\end{equation}
where we defined $x=-\log(r)$. We note that the limit $\sigma(r,t)\gg1$, which indicates that the equations are evaluated close to the origin, removes all the time derivatives and $\alpha$-dependence, making the dynamical system \eqref{linsys} formally equivalent to the one found in the static case \cite{Bonanno:2019rsq}.

\begin{figure}[htb]
\includegraphics[width=\columnwidth]{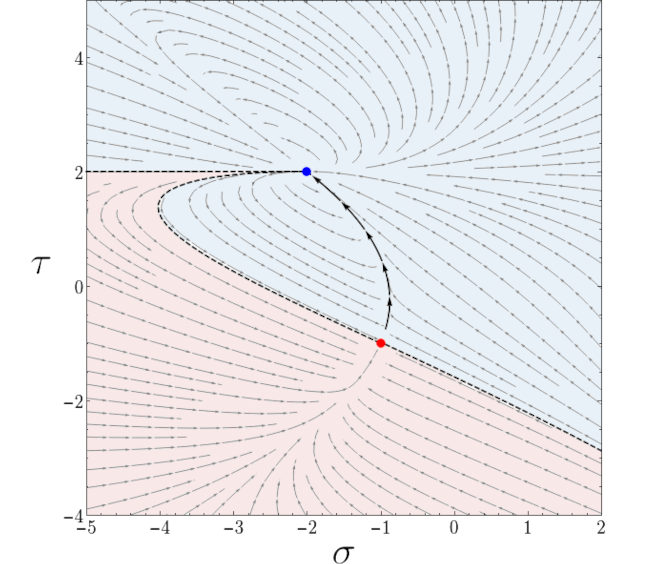}
\caption{Flow of $\tau,\ \sigma$ under the dynamical system \eqref{linsys}; the arrows point in the direction of decreasing radius. The red and blue regions indicate the points that are attracted either to $(-1,-1)$ or $(2,-2)$, and the bold arrows indicate the $(-1,-1)\to(2,-2)$ transition.}
\label{fig:origin}
\end{figure}

Figure~\ref{fig:origin} displays the flow diagram of the dynamical system in the limit $x \to \infty$, corresponding to vanishing radius. As previously established in the linear analysis of \cite{Bonanno:2019rsq}, the $(2,-2)$ fixed point exhibits stability, whereas the $(-1,-1)$ fixed point possesses only marginal stability. Crucially, incorporating time dependence uncovers an essential physical feature: the $(2,-2)$ behavior persists dynamically, while $(-1,-1)$ behavior may transition into the $(2,-2)$ configuration. This result has direct implications for Yukawa-repulsive black holes. Their behavior at the origin remains invariant under any dynamical evolution. Furthermore, since an event horizon cannot vanish instantaneously, we conclude that Yukawa-repulsive black holes represent the physically realized dynamical solution emerging from the phase transition.

To qualitatively describe the nonlinear dynamics, we analyze the weak-field limit of the time-dependent metric while maintaining asymptotic flatness. Specifically, we obtain a formal solution for $g^{rr}$ in the presence of an effective stress-energy tensor $\langle T_{\mu\nu}\rangle$:
\begin{equation}
\begin{split}
        g^{rr} \sim\, &1 - \frac{2M(t)}{r} \\
        &+ \frac{1}{2\gamma\, r}\int\mathrm{d}r\, r^2\!\!\int \mathrm{d}r'\mathrm{d}t'\, G_{(\Box-m_2^2)}\,\mathcal{C}\left(\langle T_{\mu\nu}\rangle\right),
\end{split}
\end{equation}
where $G_{(\Box-m_2^2)}$ denotes the Green's function of the Klein-Gordon operator, and $\mathcal{C}\left(\langle T_{\mu\nu}\rangle\right)$ represents a functional combination of stress-energy tensor components. This expression is physically significant because $M(t)$ asymptotically approaches the total mass as defined by both the Misner-Sharp \cite{1964PhRv..136..571M} and Hawking-Hayward \cite{Hayward:1993ph} formalisms at spatial infinity.

From the field equations and stress-energy conservation $\nabla_\mu \langle T^{\mu\nu}\rangle = 0$, we derive the large-distance behavior:
\begin{equation}\label{insta flux}
    \left(\partial_t^2 + m_2^2\right)\partial_t M(t) = \frac{1}{8\alpha}\lim_{r\to\infty}r^2\langle T_{tr}(r,t)\rangle.
\end{equation}
Remarkably, when we impose $\partial_t^3M(t) \approx 0$ (equivalent to the adiabatic approximation), we recover the standard result of General Relativity. 

The mass evolution of a black hole can be determined given an energy flux. We propose the ansatz:
\begin{equation}\label{insta set}
    \lim_{r\to\infty}r^2\langle T_{tr}(r,t)\rangle \sim \frac{\hbar}{15360 \pi M_{c}^2}\mathrm{e}^{\lambda\left(M(t)-M_{c}\right)t},
\end{equation}
where the exponential term captures the unstable behavior, the prefactor ensures recovery of the thermodynamic limit as $t \to 0$, and $\lambda\left(M(t)-M_{c}\right)$ heuristically incorporates the growing imaginary frequency \cite{Held:2022abx}. Crucially, for evolution toward Yukawa-repulsive black holes, the energy flux must be positive - opposite to the conventional case. This sign reversal has dual interpretations: at the quantum level, it corresponds to emission of ghosts with negative kinetic energy; at the classical solution level, it indicates absorption of vacuum energy by the black hole during its final phase.

\begin{figure}[htb]
\includegraphics[width=\columnwidth]{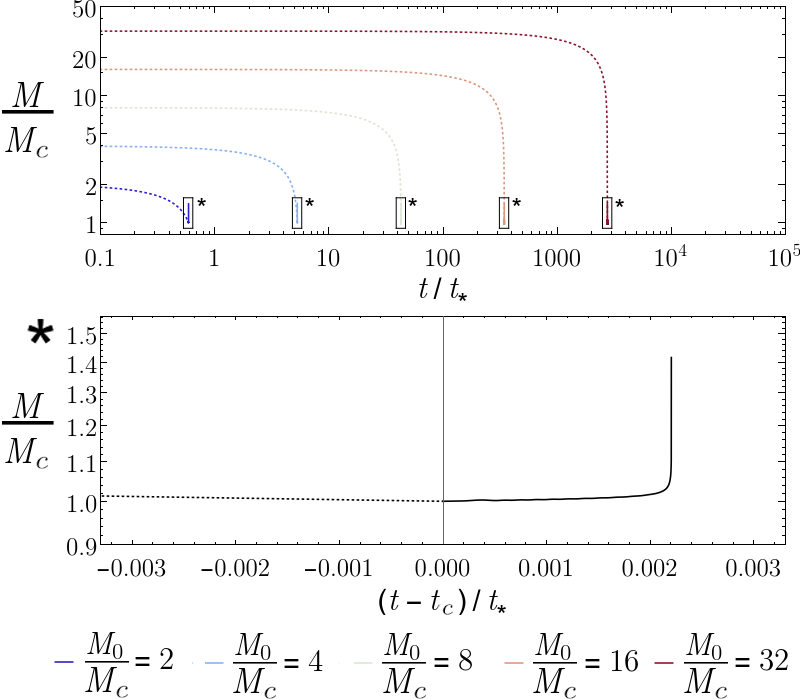}
\caption{Evaporation of Schwarzschild black holes of initial mass $M_0$ which undergo a transition into the Yukawa repulsive branch; in the bottom panel we show the common behavior close to the transition.}
\label{fig:evap}
\end{figure}

Figure~\ref{fig:evap} displays the evolution of Schwarzschild black holes with varying initial masses during ghost-induced phase transitions, using the reduced time coordinate $t_* = 10^5\alpha^{3/2}t_P$. The bottom panel specifically reveals the universal behavior near the transition point. While the energy flux in Eq.~\eqref{insta set} predictably drives extremely rapid mass growth, the evolutionary endpoint must necessarily be a stable, stationary solution. For black holes following the Yukawa-repulsive branch, this endpoint can only correspond to the limiting solution with mass $M_{\text{mtp}}$ where the horizon radius vanishes - the point at which we truncate our evolutionary calculations.

This special solution, which we previously termed the \emph{massive triple point} in \cite{Silveravalle:2022wij}, shares with Minkowski space the distinction of being a limiting case for all solution types in the parameter space. However, unlike Minkowski space (the standard evaporation endpoint in General Relativity), the massive triple point possesses both finite mass and a central singularity. Such naked singularities as evaporation endpoints are not unprecedented in modified gravity theories, having been proposed in Einstein-dilaton-Gauss-Bonnet gravity \cite{Corelli:2022pio}. We now outline its key characteristics.

Although numerical challenges complicate direct analysis, the dynamical system in Eq.~\eqref{linsys} indicates a $(2,-2)$-type behavior. This distinctive class of naked singularities, unique to quadratic gravity, has been extensively studied by Holdom and collaborators \cite{Holdom:2016nek,Holdom:2019bdv,Aydemir:2020xfd}. These solutions exhibit three defining features: first, they contain a strong curvature singularity; second, they can be reached in finite proper and asymptotic time; and third, they represent complete causal visibility, making them paradigmatic examples of naked singularities.

Remarkably, preliminary linear perturbation analysis \cite{Silveravalle:2023lnl} demonstrates their stability, solidifying the massive triple point as a physically plausible evaporation endpoint. While naked singularities present theoretical challenges, they may not conflict with observations. In particular, three crucial aspects differentiate them from their General Relativity counterparts: the gravitational potential remains universally attractive; photons emitted near the singularity experience extreme redshift (unlike the blueshift characteristic of GR naked singularities); and emission precisely at the singularity would require infinite energy to escape.

This suggests a \emph{weaker} cosmic censorship scenario: even when causally connected to our universe, the singularity may effectively hide its presence by preventing information escape. A comprehensive investigation of these naked singularities and the massive triple point will be presented in future work.

\section{Conclusions}

To conclude, we have demonstrated that incorporating general quadratic corrections to the gravitational action prevents complete evaporation of Schwarzschild black holes into Minkowski vacuum. This phenomenon stems from the presence of a ghost particle, which typically requires artificial removal in quadratic theories but emerges naturally from quantum corrections. 

The ghost particle's instability triggers a phase transition that generates a non-zero Ricci tensor and fundamentally alters the singularity's nature. This mechanism bears strong resemblance to spontaneous scalarization processes, where black holes acquire scalar charges through scalar field instabilities \cite{Doneva:2022ewd}. 

Our analysis reveals two crucial results: first, the modified singularity remains stable throughout the evolutionary process; second, this stable singularity persists as the endpoint of evaporation even after horizon disappearance. These findings have profound implications for two longstanding problems in black hole physics.  The singularity problem and information paradox become intrinsically linked in quadratic gravity. In our framework, the information paradox is naturally resolved because while the black hole evaporates, its singularity persists indefinitely. Although the information trapped within the singularity remains inaccessible, there exists no physical basis to conclude that this information is destroyed.

\begin{acknowledgments}
\section{Acknowledgments.} 
We would like to thank Aaron Held, Roman Konoplya, Andrea Spina and Alexander Zhidenko for the useful discussions. This work was partially funded by the European Union - NextGenerationEU, in the framework of the PRIN Project “Charting unexplored avenues in Dark Matter" (20224JR28W) and the INFN projects FLAG and QUAGRAP.
\end{acknowledgments}

\bibliographystyle{apsrev4-1}

\end{document}